\documentclass[12pt]{article}

\usepackage{a4}

\newlength{\abstwidth}
\def\be{\begin{equation}}
\def\ee{\end{equation}}

\begin{document}

\def\lsim{\mathrel{\rlap{\lower4pt\hbox{\hskip1pt$\sim$}}
    \raise1pt\hbox{$<$}}}         
\def\gsim{\mathrel{\rlap{\lower4pt\hbox{\hskip1pt$\sim$}}
    \raise1pt\hbox{$>$}}}         

\pagestyle{empty}

\begin{flushright}
\end{flushright}

\vspace{\fill}

\begin{center}
{\Large\bf The energy dependence of the saturation scale in DIS at low $x$$^*$}
\\[1.8ex]
{\bf Masaaki Kuroda}$^{**}$ \\[1.2mm]
Laboratoire de Physique Math\`ematique et Th\`eorique, CNRS UMR
5825 \\[1.2mm]
Universit\'e Montpellier II, Montpellier 34095, France \\[1.2mm]
and \\[1.5ex]
{\bf Dieter Schildknecht} \\[1.2mm]
Fakult\"{a}t f\"{u}r Physik, Universit\"{a}t Bielefeld \\[1.2mm]
D-33501 Bielefeld, Germany \\[1.2mm]
and \\[1.5ex]
Ludwig-Maximilians-Universit\"at M\"unchen, Sektion Physik \\[1.2mm]
D-80333 M\"unchen, Germany \\[1.5ex]
\end{center}

\vspace{\fill}

\begin{center}
{\bf Abstract}\\[2ex]
\begin{minipage}{\abstwidth}
Consistency of the previously suggested color-dipole representation of
deep-inelastic scattering (DIS) and vector-meson production 
at low $x$ with DGLAP evolution allows one to predict the exponent
of the $W^2$ dependence of the saturation scale, $\Lambda^2_{{\rm sat}} (W^2)
\sim (W^2)^{C_2}$.
One finds $C_2^{{\rm theory}} = 0.27$ in agreement with the model-independent
analysis of the experimental data from HERA on deep-inelastic electron 
scattering.  
\end{minipage}
\end{center}

\vspace{\fill}
\noindent

\rule{60mm}{0.4mm}\vspace{0.1mm}

\noindent
${}^*$ Supported by DFG-grant Schi 189/6-1\\
${}^{**}$ On leave of absence from Institute of Physics, Meiji Gakuin University,
Yokohama 244, Japan

\clearpage
\pagestyle{plain}
\setcounter{page}{1}


The present paper is concerned with deep inelastic electron scattering (DIS) at
low $x \cong Q^2/W^2 \ll 1$. In short, we analyse the consistency between 
DGLAP evolution of the nucleon structure function $F_2 (x , Q^2)$  and the
color-dipole picture. We find that the exponent $C_2^{{\rm theory}}$  
that in our formulation of the color-dipole approach determines the 
energy dependence of the total photoabsorption cross section at large
$Q^2$, $\sigma_{\gamma^* p} (W^2 , Q^2) \sim (W^2)^{C_2} / Q^2$, or, 
equivalently, the energy dependence of the ``saturation scale'' 
$\Lambda^2_{{\rm sat}} (W^2) \sim (W^2)^
{C_2}$, coincides with the result of previous fits to the experimental data,
$C_2^{{\rm theory}} \cong C_2^{{\rm experiment}}$. 

For $x \cong Q^2/W^2 \ll 1$,
the photon-proton interaction is dominated by the interaction of the photon
with the quark-antiquark sea in the proton. The proton structure 
function for $x\ll 1$ only contains the flavor-singlet quark distributions, 
and their evolution in $Q^2$ for $Q^2 \ge Q^2_0$ \cite{1} is in good 
approximation determined by the gluon structure function alone \cite{2},
\be
\frac{\partial F_2 \left( \frac{x}{2} , Q^2 \right)}{\partial \ln Q^2} =
\frac{R_{e^+ e^-}}{9\pi} \alpha_s(Q^2) x g (x , Q^2) . 
\label{(1)}
\ee
The notation in (\ref{(1)}) is the standard one, we only note 
$R_{e^+e^-} = 3 \sum_f Q^2_f=10/3$, where $Q_f$ denotes the quark charge and $f$
runs over the contributing $(n_F = 4)$ flavors. In the physical picture 
underlying (\ref{(1)}) the 
perturbatively calculable photon-gluon-scattering amplitude is
supplemented by the gluon structure function that parameterizes the 
unknown proton properties to be determined experimentally.
The structure function $F_2 (x,Q^2)$ is related to the flavor-blind,
flavor-singlet quark distribution
\be
x\Sigma (x, Q^2) = n_F (x q (x, Q^2) + x \bar q (x , Q^2))
\label{(2)}
\ee
via
\be
F_2 (x, Q^2) = \frac{R_{e^+ e^-}}{12} x \Sigma (x, Q^2) .
\label{(3)}
\ee
 
In the color-dipole picture \cite{3}, valid at low $x \ll 1$ and any $Q^2 \ge
0$, in terms of the imaginary part of the virtual-photon-proton 
forward-scattering amplitude, the process of $\gamma^* p$ scattering 
proceeds via the fluctuation of the photon into a $q \bar q$ pair that
subsequently scatters on the proton via (the generic structure of) two-gluon
exchange \cite{4}. 
The properties of the proton are contained in the color-dipole cross section
\be
\sigma_{(q \bar q)p} ( \vec r_\bot , W^2) = \int d^2 l_\bot
\tilde\sigma_{(q \bar q)p} (\vec l_\bot^{~2} , W^2 ) (1 - e^{-i\vec l_\bot \vec r_
\bot} ) 
\label{(4)}
\ee
that depends on the two-dimensional transverse quark-antiquark separation,
$\vec r_\bot$, \cite{5,3}.
The function
$\tilde\sigma (\vec l^{~2}_\bot , W^2)$ is associated with the 
gluon-transverse-momentum distribution in the proton, and the factor 
$(1 - \exp (-i \vec l_\bot 
\vec r_\bot))$ in (\ref{(2)}) is characteristic of the (QCD) gauge-theory 
structure. This factor originates from the couplings of the two gluons to 
either the same quark (antiquark) or to a quark and an antiquark. Motivated by
the mass dispersion relation of generalized vector dominance \cite{6,7} or,
equivalently, life-time arguments \cite{8} on the hadronic (quark-antiquark)
fluctuation of the virtual photon, the energy $W$ appears as the second 
variable besides $\vec r_\bot$ or $\vec l_\bot$ in (\ref{(4)}). We refer to the
literature \cite{3,5} for the explicit representation of the total 
virtual photoabsorption
cross section in terms of the virtual-photon wave function describing the $q \bar
q$ fluctuations of the photon and the dipole cross section in (\ref{(4)}). 

At sufficiently large $Q^2$, the dipole cross section in the limit of 
small interquark transverse separation, $\vec r^{~2}_\bot \rightarrow 0$, 
becomes relevant. From (\ref{(4)}), for $\vec r^{~2}_\bot \rightarrow 0$,
\footnote{Note that the scale for $\vec r_\bot^{~2}$ that determines the 
validity of the expansion in (\ref{(5)}) depends on the behavior of 
$\tilde\sigma (\vec l_\bot^{~2} , W^2)$. Essentially, it is given by the
effective or average value of $\vec l_\bot^{~2}$ determined by 
$\tilde\sigma (\vec l_\bot^{~2} , W^2)$.}
\be
\sigma_{(q \bar q)p} (\vec r_\bot , W^2) \cong \frac{1}{4} \vec r_\bot^{~2} \pi
\int d \vec l_\bot^{~2} \vec l_\bot^{~2} \tilde\sigma_{(q \bar q)p} 
(\vec l_\bot^{~2} , W^2).
\label{(5)}
\ee
By reformulating the $\gamma^*$-gluon-scattering approach underlying (\ref{(1)})
in terms of the transverse position-space variable $\vec r_\bot$, one finds
\cite{9,10} 
that the gluon structure function $x g (x , Q^2)$ in (\ref{(1)}) is 
proportional to the right-hand side of (\ref{(5)}),
\be
    \alpha_s(Q^2) xg (x, Q^2) = \frac{3}{4\pi} \int d \vec l_\bot^{~2} 
    \vec l_\bot^{~2} \tilde\sigma_{(q \bar q)p} (\vec l_\bot^{~2} , W^2).
\label{(6)}
\ee
The gluon structure function, according to (\ref{(6)}), is proportional 
to the first moment of the gluon transverse-momentum distribution.\footnote{
Note that the $Q^2$ dependence of the gluon structure function at fixed $x$
is contained in $W^2 = Q^2 / x$. This is at variance with the conventional 
assumption, where $x$ occurs on the right-hand side in (\ref{(6)}) and the $Q^2$ 
dependence is intro6uced via the upper limit, $Q^2$, of the integral in 
(\ref{(6)}).} 
The 
validity of (\ref{(6)}) is restricted to sufficiently large $Q^2$, where both
the notion of the gluon-structure function appearing in (\ref{(1)}) as well as
the $\vec r_\bot^{~2} \rightarrow 0$ expansion of the dipole cross section in 
(\ref{(5)}) are applicable. 

Actually only transverse and longitudinal $(q \bar q)^{J=1}_{T,L}$ 
(vector) states 
contribute to the imaginary part of the virtual forward-scattering Compton 
amplitude. The structure function $F_2 (x,Q^2)$ may be represented \cite{11}
in terms of the $J=1$ projections of the color-dipole cross section (\ref{(4)}).
The leading contribution to 
\be
F_2 (x , Q^2) = \frac{Q^2}{4\pi^2\alpha} \sigma_{\gamma^* p} (W^2 , Q^2)
= \frac{Q^2}{4\pi^2\alpha} (\sigma_{\gamma^*_T p} (W^2 , Q^2) + 
\sigma_{\gamma^*_L p} (W^2 , Q^2))
\label{(7)}
\ee
at sufficiently large $Q^2$ becomes\footnote{The expression (8) for $F_2
(x, Q^2)$ is obtained from (3.13) and (3.14) in ref.\cite{11} by expansion in
powers of $\vec l_\bot^{~\prime 2} / (4 Q^2 + \vec l_\bot^{~\prime 2})$.}
\begin{eqnarray}
F_2 (x , Q^2) & = & 
\frac{Q^2}{36\pi^2} R_{e^+ e^-} \left( \int d \vec l^{~\prime 2}_\bot 
\frac{4\vec l^{~\prime 2}_\bot}{4 Q^2 + \vec l^{~\prime 2}_\bot} \bar\sigma_
{(q \bar q)^{J=1}_{T}} (\vec l^{~\prime 2}_\bot , W^2)\right.  \nonumber \\
& & \left. + \frac{1}{2} \int d \vec l^{~\prime 2}_\bot \frac{4 
\vec l^{~\prime 2}_\bot}
{4 Q^2 + \vec l^{~\prime 2}_\bot} \bar\sigma_{(q \bar q)^{J=1}_{L}} ( \vec l^
{~\prime 2}_\bot , W^2) \right) \nonumber \\
& \cong & \frac{R_{e^+ e^-}}{36 \pi^2} \left( \int d \vec l^{~\prime 2}_\bot 
\vec l^{~\prime 2}_\bot  \bar\sigma_{(q \bar q)^{J=1}_{T}} (\vec l^{~\prime 2}_
\bot , W^2)\right. \label{(8)} \\
& +& \left. \frac{1}{2} \int d \vec l^{~\prime 2}  \vec l^{~\prime 2} 
\bar\sigma_{(q \bar q)^{J=1}_{L}} ( \vec l^{~\prime 2} , W^2)\right) ,\nonumber 
\end{eqnarray} 
where $\vec l^{~\prime 2}$ is related to the gluon transverse momentum and 
the light-cone variable $z$ via
\be
\vec l^{~\prime 2} = \frac{\vec l^{~2}_\bot}{z (1-z)} . 
\label{(9)}
\ee
Moreover, also the gluon structure function (\ref{(6)}) may be represented in 
terms of 
the longitudinal part of the $J=1$ projection of the color-dipole cross section, 
\cite{11},
\be
\alpha_s (Q^2) x g (x, Q^2) = \frac{1}{8\pi} \int d \vec l^{~\prime 2}_\bot 
\vec l^{~\prime 2}_\bot \bar\sigma_{(q \bar q)^{J=1}_{L}} (\vec l^{~\prime 2}_\bot
, W^2).
\label{(10)}
\ee
In terms of the ``saturation scale''
\be
\Lambda^2_{{\rm sat}} (W^2) \equiv \frac{\pi}{\sigma^{(\infty)}} \int
d \vec l^{~\prime 2}_\bot \vec l^{~\prime 2}_\bot \bar\sigma_{(q \bar q)^{J=1}_
{L}} (\vec l^{~\prime 2}_\bot , W^2), 
\label{(11)}
\ee
where the constant $\sigma^{(\infty )}$ will be explicitly defined below
(compare (\ref{(40)})), 
(\ref{(10)}) becomes 
\be
\alpha_s (Q^2) x g (x , Q^2) = \frac{1}{8\pi^2} \sigma^{(\infty)} \Lambda^2_
{{\rm sat}} (W^2).
\label{(12)}
\ee

So far our considerations have exclusively been based on the two-gluon 
exchange structure embodied in the form of the color-dipole cross section 
(\ref{(4)}). 
To proceed, we assume the flavor-singlet distribution (\ref{(2)}) and 
the gluon distribution (\ref{(12)}) to have identical dependence 
on the kinematic 
variables $x$ and $Q^2$. In our case, $x$ and $Q^2$ appear in the combination 
$W^2 \cong Q^2/x$. Both $x \Sigma (x , Q^2)$ and $\alpha_s (Q^2) x g (x, Q^2)$
must then be proportional to $\Lambda^2_{{\rm sat}}(W^2)$. Since $F_2 (x , Q^2)$
is proportional to $x \Sigma (x , Q^2)$, compare (\ref{(3)}), also $F_2 (x, Q^2)$
in (\ref{(8)}) must be proportional to $\Lambda^2_{{\rm sat}} (W^2)$.

Since, moreover, the longitudinal term on the right-hand side in (\ref{(8)})
is proportional to the gluon structure in (\ref{(10)}) and (\ref{(12)}), also
the transverse contribution to $F_2 (x, Q^2)$ in (\ref{(8)}) must be proportional
to $\Lambda^2_{{\rm sat}}(W^2)$. In terms of the integrals in (\ref{(8)}), the 
above requirement on the flavor singlet quark and the gluon distribution thus 
becomes 
\be
\int d \vec l^{~\prime 2}_\bot \vec l^{~\prime 2}_\bot \bar\sigma_{(q \bar q)^
{J=1}_{T}} (\vec l^{~\prime 2}_\bot , W^2) = r \int d \vec l^{~\prime 2}_\bot
\vec l^{~\prime 2}_\bot \bar\sigma_{(q \bar q)^{J=1}_{L}} (\vec l^{~\prime 2}_\bot
, W^2).
\label{(13)}
\ee
Note that the constant $r$ is related to the longitudinal to transverse 
ratio, 
\be
\frac{\sigma_{\gamma^*_L p} (W^2 , Q^2)}{\sigma_{\gamma^*_T p} (W^2 , Q^2)}
= \frac{1}{2 r} .
\label{(14)}
\ee
Our previous analysis \cite{12,13,14,15} of the experimental data on DIS was based
on the equality of
\be
\bar\sigma_{(q \bar q)^{J=1}_{T}} (\vec l^{~\prime 2}_\bot , W^2) = 
\bar\sigma_{(q \bar q)^{J=1}_{L}} (\vec l^{~\prime 2}_\bot , W^2) , 
\label{(15)}
\ee
i.e. on
\be 
r = 1 . 
\label{(16)}
\ee
We found consistency with the experimental data, including \cite{14} the 
available information on the longitudinal virtual photoabsorption cross section.
We will
henceforth put $r=1$. Upon inserting (\ref{(13)}) and intoducing 
$\Lambda^2_{{\rm sat}} (W^2)$ from (\ref{(11)}), $F_2 (x , Q^2)$ from 
(\ref{(8)}) becomes 
\be F_2 (x , Q^2) = \frac{R_{e^+ e^-}}{36\pi^3} \sigma^{(\infty)} \Lambda^2_
{{\rm sat}}(W^2) \left( 1 + \frac{1}{2} \right) , 
\label{(17)}
\ee
where the sum on the right-hand side refers to the sum of the transverse and 
longitudinal parts. 

We now insert $F_2 (x, Q^2)$ from (\ref{(17)}) and the gluon distribution 
(\ref{(12)}) into the DGLAP-evolution equation (\ref{(1)}), to find the 
interesting constraint
\be
\frac{\partial}{\partial \ln W^2} \Lambda^2_{{\rm sat}} (2 W^2) = \frac{1}{3}
\Lambda^2_{{\rm sat}} (W^2), 
\label{(18)}
\ee
or, alternatively, in terms of the observable $F_2 (x , Q^2)$, 
\be
\frac{\partial}{\partial \ln W^2} F_2 (2 W^2) = \frac{1}{3} F_2 (W^2) . 
\label{(19)}
\ee 

We adopt a power-law ansatz for $\Lambda^2_{{\rm sat}} (W^2)$, 
\be
\Lambda^2_{{\rm sat}} (W^2) = B \left( \frac{W^2}{W^2_0} + 1 \right)^{C_2} 
\cong B \left(\frac{W^2}{W^2_0} \right)^{C_2} , 
\label{(20)}
\ee
identical in form to the one previously employed \cite{12,13,14,15} in 
(successful) fits to the experimental data on the virtual 
photoabsorption cross section,
\be
\sigma_{\gamma^* p} (W^2 , Q^2) = \sigma_{\gamma^* p} (\eta (W^2, Q^2)),
\label{(21)}
\ee
and on deeply virtual Compton scattering (DVCS) 
in terms of the scaling variable \cite{12,13}\footnote{For the sake of clarity, 
we 
introduced the notation $\Lambda^2_{{\rm sat}} (W^2) \equiv \Lambda^2 (W^2)$
for the quantity previously denoted by $\Lambda^2 (W^2)$.}
\be
\eta (W^2, Q^2)\equiv \frac{Q^2 + m^2_0}{\Lambda^2_{{\rm sat}} (W^2)} . 
\label{(22)}
\ee
Inserting the power-law ansatz (\ref{(20)}) into (\ref{(18)}), we deduce
\be
C_2 = \frac{1}{3} \left( \frac{1}{2} \right)^{C_2} ,  
\label{(23)}
\ee
and accordingly\footnote{The arguments leading to (\ref{(24)}) were
implicitly used in the third paper of ref.\cite{13} without,
however, fully realizing their significance.}
\be
C_2^{{\rm theory}} = 0.276 . 
\label{(24)}
\ee

The theoretically deduced magnitude of the exponent 
$C_2^{{\rm theory}} = 0.276$ in (\ref{(24)}) is in agreement with the 
model-independent fit \cite{12,13} to the experimental data, 
\be
C_2^{\exp} = 0.275 \pm 0.06 . 
\label{(25)}
\ee
The model-independent fit assumes that $\sigma_{\gamma^* p} (W^2 , Q^2)$ may 
be represented by a smooth function of the scaling variable $\eta (W^2, Q^2)$
in (\ref{(22)}) with $\Lambda^2_{{\rm sat}} (W^2)$ from (\ref{(20)}). As a 
consequence of the generality of the ansatz, any theoretical prejudice with 
respect to the empirical validity of scaling in $\eta (W^2, Q^2)$ is excluded.

The fit based on an explicit ansatz for the color-dipole cross section 
(compare (\ref{(32)}) below) gave the more precise result
\be
C_2^{{\rm exp}} = 0.27 \pm 0.01
\label{(26)}
\ee
in agreement with our theoretical result (\ref{(24)}).

To summarize: the choice of $W$ as the relevant variable in the 
color-dipole approach together with the requirement that the singlet 
quark distribution and the gluon distribution have identical dependence on the 
kinematic variables, $W^2 = Q^2 / x$ in our case, converts the 
DGLAP-evolution equation (\ref{(1)}) into a constraint that allows one to 
predict the exponent $C_2$. The agreement with experiment supports the 
validity, at least as a relevant approximation, of the underlying assumptions.

The choice of $W^2 \cong Q^2 / x$ as the relevant variable in our approach 
was motivated by the color-dipole approach and its generalized-vector-dominance
interpretation. We note that the dependence (\ref{(17)})
\be
F_2 \sim (W^2)^{C_2}
\label{(27)}
\ee
is closely related to the so-called singular solution of the gluon evolution 
equation, where\footnote{Compare the discussion in section 7 of ref.\cite{16} 
and the literature quoted there, in particular ref.\cite{17}.}
\be
F_2 \sim \ln Q^2 x^{-\lambda} \simeq \frac{\ln Q^2}{(Q^2)^\lambda} (W^2)^
\lambda \cong (W^2)^\lambda 
\label{(28)}
\ee
with $\lambda \ge 0.25$ being fixed and equal to the input value at all
$Q^2$. In a restricted but relevant range of $Q^2$, (\ref{(27)}) is similar to 
(\ref{(28)}). 

The conventional application of the color-dipole approach \cite{3} to DIS
does not explicitly introduce the $J=1$ projection of the color-dipole 
forward-scattering amplitude. One starts by an assumption on $\tilde\sigma
(\vec l^{~ 2}_\bot , W^2)$ in (\ref{(4)}), rather than its $J=1$
projection, $\bar\sigma_{(q \bar q)^{J=1}_{T,L}} (\vec l^{~\prime 2}_\bot , 
W^2)$. We briefly elaborate on how our approach can be formulated in terms
of $\tilde\sigma (\vec l^{~ 2}_\bot , W^2)$.

Consider the ansatz
\be
\tilde\sigma (\vec l^{~ 2}_\bot , W^2) = \frac{\sigma^{(\infty)}}{\pi}
\delta \left( \vec l^{~ 2}_\bot - \frac{1}{6} \Lambda^2_{{\rm sat}}
(W^2) \right). 
\label{(29)}
\ee
Inserting (\ref{(29)}) into (\ref{(6)}), we immediately recover the gluon 
structure function (\ref{(12)}).
Evaluation of the $J=1$ parts of (\ref{(29)}) yields 
\be
\bar\sigma_{(q \bar q)^{J=1}_{L,T}} (\vec l^{~\prime 2}_\bot , W^2) = f_{L,T}
(\vec l^{~\prime 2}_\bot, \Lambda^2_{{\rm sat}}(W^2)) \theta (
\vec l^{~\prime 2}_\bot - \frac{2}{3} \Lambda^2_{{\rm sat}} (W^2)) , 
\label{(30)}
\ee
the explicit form of the function $f_{L,T} (\vec l^{~\prime 2}_\bot , 
\Lambda^2_{{\rm sat}} (W^2))$ being irrelevant in the present context. We only 
note the normalization of (\ref{(30)})
\be
\int d \vec l^{~\prime 2}_\bot \bar\sigma_{(q \bar q)^{J=1}_{T}} 
(\vec l^{~\prime 2}_\bot , W^2) = \int d \vec l^{~\prime 2}_\bot 
\bar\sigma_{(q \bar q)^{J=1}_{L}} (\vec l^{~\prime 2}_\bot , W^2) = 
\frac{\sigma^{(\infty)}}{\pi} . 
\label{(31)}
\ee
Evaluating the longitudinal part of $F_2 (x,Q^2)$ in (\ref{(8)}) by substituting 
(\ref{(30)}), we recover our previous result (\ref{(17)}) for the 
longitudinal contribution. The normalization (\ref{(31)}) suggests to 
approximate (\ref{(30)}) by
\be
\bar\sigma_{(q \bar q)^{J=1}_{T}} (\vec l^{~\prime 2}_\bot , W^2) = 
\bar\sigma_{(q \bar q)^{J=1}_{L}} (\vec l^{~\prime 2}_\bot , W^2) = 
\frac{\sigma^{(\infty)}}{\pi} \delta (\vec l^{~\prime 2}_\bot - \Lambda^2_
{{\rm sat}}(W^2)). 
\label{(32)}
\ee
With (\ref{(32)}) inserted into (\ref{(8)}), we now obtain not only the 
longitudinal, but also the transverse part of $F_2$ given in (\ref{(17)}).
The direct evaluation of $F_2$, inserting (\ref{(30)}), however, yields 
\be 
F_2 (x , Q^2) = \frac{R_{e^+ e^-}}{36\pi^3} \sigma^{(\infty)} \Lambda^2_
{{\rm sat}}(W^2) \left( \frac{1}{2} \ln \frac{6 Q^2}{\Lambda^2_{{\rm sat}} (W^2)}
+ \frac{1}{2} \right). 
\label{(33)}
\ee
in distinction from (\ref{(17)}). Our requirement of identical singlet 
quark and gluon distributions that is contained in (\ref{(13)}),
(\ref{(15)}) and (\ref{(32)}) is not 
fulfilled by the ansatz (\ref{(29)}). It is not known whether an ansatz for 
$\tilde\sigma (\vec l^{~ 2}_\bot , W^2)$ can be given such that (\ref{(13)})
with $r=1$
be valid. For the time being, we have to accept the equality (\ref{(32)}) as 
a valid approximation for the $J=1$ projections that describes the 
experimental data on DIS in the low $x$
diffraction region with the predicted value of the exponent $C_2$.  

Encouraged by the result (\ref{(24)}), we now examine the coupled system of 
equations for singlet quark and gluon evolution.

Substituting the gluon structure function (\ref{(12)}) and the singlet quark 
distribution
$x\Sigma (x, Q^2)$ from (\ref{(3)}) with (\ref{(17)}), 
into the evolution equations, with the power-law (\ref{(20)}) for 
$\Lambda^2_{{\rm sat}} (W^2)$ 
and the notation $t \equiv \ln Q^2$, one finds
\be
\frac{\partial \Lambda^2_{{\rm sat}} (W^2)}{\partial \ln W^2} = 
\Lambda^2_{{\rm sat}} (W^2) \int^1_x
dy \left(\frac{\alpha_s (t)}{2\pi} P_{q q} (y) + P_{q g} (y) \right) y^{C_2} , 
\label{(34)}
\ee
and
\begin{eqnarray}
\frac{\partial \Lambda^2_{{\rm sat}} (W^2)}{\partial \ln W^2} & = &
\Lambda^2_{{\rm sat}} (W^2) 
\frac{1}{\alpha_s (t)} \frac{d\alpha_s (t)}{dt} + \nonumber \\
& + &\Lambda^2_{{\rm sat}} (W^2) \frac{\alpha_s (t)}{2\pi} \int^1_x dy \left( 
\frac{\alpha_s (t) n_f}{\pi} P_{gq} (y) + P_{gg} (y) \right) y^{C_2} .
\label{(35)}
\end{eqnarray}
The first equation, (\ref{(34)}), without relying on the approximation contained 
in the right-hand side of (\ref{(1)}),  
describes the
evolution of the flavor singlet quark distribution, while the second equation, 
(\ref{(35)}), describes the evolution of the gluon distribution. By noting 
that $\partial \Lambda^2_{{\rm sat}} (W^2) / \partial \ln W^2 = C_2 
\Lambda^2_{{\rm sat}} (W^2)$, 
and upon evaluating the integrals on the right hand side, in (\ref{(34)}), 
we obtain 
\be
   C_2 =   0.044 \alpha_s (t)
         +\frac{C^2_2 + 3 C_2 + 4}{2 (C_2 + 1)(C_2 + 2) (C_2 + 3)}.
\label{(36)}
\ee
The numerical value of $0.044$ in the (small) $C_2$-dependent correction 
proportional to 
$\alpha_s (t)$ in 
(\ref{(36)}) was obtained by inserting $C_2 = 0.276$.
Solving (\ref{(36)}) for $C_2$, we find  
\begin{eqnarray}
C_2 & = &  0.044 \alpha_s (t)  + 0.260\nonumber \\
  & \cong & 0.265 , 
\label{(37)}
\end{eqnarray}
upon disregarding the (weak) $Q^2$ dependence of $\alpha_s$  at large
$Q^2$ and inserting $\alpha_s = 0.11$. 

A similar approach, when applied to the gluon-evolution equation 
(\ref{(35)}), 
leads to 
\begin{eqnarray}
C_2 & = & \frac{1}{\alpha_s (t)} \frac{d \alpha_s (t)}{dt} + C_3 (C_2) 
\alpha^2_s (t) + C_4 (C_2) \alpha_s (t) \nonumber \\
 & \cong & 0.275 
\label{(38)}
\end{eqnarray}
The dependence of the coefficients $C_3 (C_2)$ and $C_4 (C_2)$ on $C_2$ 
is directly calculated from (\ref{(35)}).
The value of $C_2 = 0.275$ was obtained by consistently solving 
(\ref{(38)}), using $\alpha_s (t) = 0.11$.

As a result of our analysis of the complete evolution equations,
by comparing (\ref{(37)}) and (\ref{(38)}) with (\ref{(24)}), we
conclude that the power-law ansatz (\ref{(20)}) with a constant 
($Q^2$-independent) value of $C_2^{{\rm theory}}$ of magnitude 
$C_2^{{\rm theory}} \cong 0.276$ according to (\ref{(24)})
is consistent with evolution.
The additional $\alpha_s$-dependent contributions 
in (\ref{(34)}) which have been 
ignored in (\ref{(1)}) 
hardly affect the value of the exponent $C_2$. 

So far in this paper we were concerned with DIS at low $x \ll 1$ and 
sufficiently large $Q^2 \gg \Lambda^2_{{\rm sat}} (W^2)$ 
where the QCD-improved parton picture and the 
color-dipole picture are dual descriptions of the underlying physics. 

For details on the extension to $Q^2 \ll \Lambda^2_{{\rm sat}} (W^2)$
and the (successful) description of the experimental data based on the ansatz
(\ref{(32)}), we refer to 
refs.\cite{13,14,15}. 
We only note the $Q^2 \rightarrow 0$ limit in addition to the large-$Q^2$
limit,  
\begin{eqnarray}
\sigma_{\gamma^* p} (W^2 , Q^2) & = & \sigma_{\gamma^* p} (\eta (W^2, Q^2)) 
= \nonumber\\
& = & \frac{\alpha}{3\pi} R_{e^+ e^-} \sigma^{(\infty)} \left\{
\matrix{ \ln \frac{\Lambda^2_{{\rm sat}}(W^2)}{Q^2 + m^2_0} , & (Q^2\ll 
\Lambda^2_{{\rm sat}} (W^2)) , \cr
\frac{\Lambda^2_{{\rm sat}}(W^2)}{2 Q^2} , 
& (Q^2 \gg \Lambda^2_{{\rm sat}} (W^2)) . \cr} \right. 
\label{(39)}
\end{eqnarray}
With $\Lambda^2_{{\rm sat}}(W^2)$ from (\ref{(11)}),
the asymptotic limit of $Q^2 \gg \Lambda^2_{{\rm sat}} (W^2)$, 
in (\ref{(39)}) coincides with (\ref{(17)}). 
We also note 
\be
   \sigma^{(\infty)} = \pi \int d \vec l_\bot^{~\prime 2} 
    \bar\sigma_{(q \bar q)^{J=1}_L p} (\vec l_\bot^{~\prime 2} , W^2) . 
\label{(40)}
\ee
Hadronic unitarity requires $\sigma^{(\infty)}$ to at most show 
a weak $W$ dependence. The fit to the experimental data led to 
$\sigma^{(\infty)} = $ constant. 

It is worth noting that the virtual photoabsorption cross section, 
or $F_2 (x , Q^2)$,
for $x = Q^2 / W^2 \ll 1$ and any $Q^2$ only depends on the integrated 
quantities in (\ref{(11)}) and (\ref{(40)}).

From (\ref{(39)}), at any $Q^2$, for sufficiently large energy, such that
$\Lambda^2_{{\rm sat}} (W^2) \gg Q^2$ and $\eta (W^2, Q^2) \rightarrow 0$, 
the virtual photoabsorption cross section approaches the ``saturation limit'' of
\cite{12,13} 
\be
\lim_{{W^2 \rightarrow \infty}\atop{{Q^2 {{\rm fixed}}}}}
\frac{\sigma_{\gamma^* p} (\eta (W^2 , Q^2))}{\sigma_{\gamma p} (W^2)} = 1.
\label{(41)}
\ee
The quantity $\Lambda^2_{{\rm sat}} (W^2)$ indeed sets the 
scale for the limiting behavior (\ref{(41)}). 
The terminology ``saturation scale'' for the effective gluon transverse 
momentum
squared, $(1/6) \Lambda^2_{{\rm sat}} (W^2)$, originating from the underlying 
two-gluon exchange, is indeed appropriate.

In conclusion: the previously formulated color-dipole approach to DIS
(the generalized vector dominance/color dipole picture, GVD-CDP)
has been examined with respect to the underlying singlet quark
and gluon distribution. Both of these distributions being proportional 
to the saturation scale, $\Lambda^2_{{\rm sat}} (W^2)$, we find that the 
evolution equations lead to a remarkable constraint on the value of the exponent 
of the $W^2$ dependence that agrees with the experimental result.

\section*{Acknowledgements}
It is a pleasure to thank Bronislav G. Zakharov for valuable discussions.

\end{document}